\documentclass[print,showpacs,twocolumn,prl]{revtex4}%
\usepackage{amsfonts}
\usepackage{amsmath}
\usepackage{amssymb}
\usepackage{graphicx}%
\begin{document}
\title{Ring-Like Solitons in Plasmonic Fiber Waveguide Composed of Metal-Dielectric Multilayers}
\author{Jie-Yun Yan$^{1,}$*, Lu Li,$^{2}$ and Jinghua Xiao$^{1}$}
\affiliation{$^{1}$School of Science, Beijing University of Posts
and Telecommunications, Beijing 100876, People's Republic of China}
\affiliation{$^{2}$Institute of Theoretical Physics, Shanxi
University, Taiyuan 030006, People's Republic of China}



\begin{abstract}
We design a plasmonic fiber waveguide (PFW) composed of coaxial
cylindrical metal-dielectric multilayers in nanoscale, and
constitute the corresponding dynamical equations describing the
modes of propagation in the PFW with the Kerr nonlinearity in the
dielectric layers. The physics is connected to the discrete matrix
nonlinear Schr\"{o}dinger equations, from which the highly confined
ring-like solitons in scale of subwavelength are found both for the
visible light and the near-infrared light in the self-defocusing
condition. Moreover, the confinement could be further improved when
increasing the intensity of the input light due to the cylindrical
symmetry of the PFW, which means both the width and the radius of
the ring are reduced.
\end{abstract}
\maketitle

How to control the propagation of the light is a most important
subject in optics. Using the technology of the optical fiber
waveguides (OFWs) to pilot the light is a big advance towards
all-optical signal processing. When consider ever-accelerated
miniaturization of optical devices, however, the conventional OFWs
seem to be difficult to fulfill the requirement because of the
diffraction limitation for the optical components of dielectric
photonic materials. Recently, it is shown that the limitation may be
overcome in the rapid developing field of
plasmonics\cite{Maier2001,Maier2005,Gramotnev2010} based on
properties of the surface plasmon polariton (SPP), a confined mode
localized at the interface of metal and dielectric material
\cite{Pitarke2007,Barnes2003}. Subsequently wide attentions have
been paid to miscellaneous nanostructures with metamaterials
involved in pursuit of subwavelength confinement of the
light\cite{Lee2010}. Among these progresses, several researchers
have predicted the subwavelength control of the light in various
lattices through the formation of solitons when the nonlinearity is
considered. It has been found that the planar nonlinear
metal-dielectric multilayers (MDM), a stack of alternating
metal-dielectric nanolayers, can effectively manipulate the
propagation of light\cite{Husakou2007,Vukovic2011,Lin2006} and put a
subwavelength confinement on it
\cite{Zhang2007,Ye2010,Peleg2009,Avrutsky2007,Zhang2009,LatticeSoliton}.
For instance, Zhang's group theoretically found the subwavelength
discrete soliton in the nanoscaled periodic structures consisting of
MDM\cite{Zhang2007}. Ye \textit{et al.} predicted the stable
fundamental and vortical plasmonic lattice solitons of subwavelength
extent in arrays of metallic nanowires embedded in a nonlinear
medium\cite{Ye2010}.

Then could we construct a kind of waveguide in analogy to the
optical fiber waveguide to control the light's propagation in
two-dimensional (2D) transverse space based on the plasmonics rather
than the optics? In this Letter we design a plasmonic fiber
waveguide (PFW) by rolling the MDM to a cylindrical shape to guide
the light. The nonlinearity in the dielectric layers are employed to
realize the subwavelength confinement in radial direction when
propagating along the axial direction. Based on the advanced
nano-technologies\cite{Jiang2004,Jzksic2011,Fedutik2007} in addition
to the sophisticated fiber fabrications, this kind of PFW is
supposed to be a easily-fabricated structures. Actually a spherical
hyperlen made of the MDM has been realized in
experiment\cite{Rho2010}. Besides, in contrast to the lattices made
of planar MDM, the PFW has the cylindrical symmetry, which produces
the centripetal confinement in the form of solitons. Our results
show that the energy could be highly confined in the cross section
with size far small than the operating wavelength. This kind of
confinement combines the advantages of fiber's long distance
transportation and minimized transverse space caused by the
cooperation of nonlinearity and SPP's field enhancement effect,
which is significant for applications of plasmonics in future
communication and large-scale integration.

\begin{figure}[ptb]
\includegraphics[width=8cm]{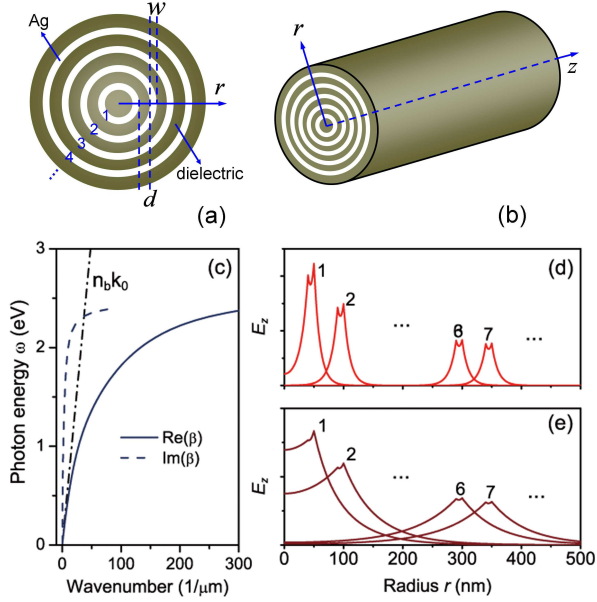}
\caption{(a) Cross section of the PFW; (b) Schematics of the PFW;
Only first few metallic layers are draw for demonstration. (c)
Dispersion of SPP mode in the 1-th SPW. $E_{z}$ distributions in
different SPWs in case of $\lambda=700$ nm (d), and $\lambda=1550$
nm (e), respectively.
For clarity, only that in SPWs with indexes 1, 2, 6, and 7 are draw.}%
\label{figure1}%
\end{figure}

Our structure of the PFW is composed of alternating metal and
dielectric coaxial cylindrical layers, as shown in Fig. 1(a) and
(b), where the coaxial cylindrical metal layers are labeled as
$n=1,2,\cdots$ along radial direction. The thicknesses of the metal
layer and the dielectric layer are represented by $w$ and $d$,
respectively. Note that the dielectric material at center is a
cylinder with the radius of $d$. The dielectric constants of metal
and dielectric material are $\epsilon _{0}\epsilon_{m}$ and
$\epsilon_{0}\epsilon_{d}$, respectively, with $\epsilon_{0}$ is
dielectric constant of the vacuum. A set of typical parameters used
in the Letter are that: the metallic dielectric constant is
Johnson's data for the metal Ag \cite{Johnson}, the dielectric
constant for the dielectric material is 10, and their thicknesses
are $w=10$nm and $d=40$nm, respectively. In fact, the following
theory is still applicable when the parameters $d$ and $w$ are
functions of $n$.

When the light wave with the frequency $\omega$ propagates in the
PFW, the SPP modes are excited at each interface\cite{Smith2010}.
Therefore the PFW can be considered as an plasmonic multiwaveguide
system, of which the $i$-th single plasmonic waveguide (SPW) is
constituted by the corresponding $i$-th coaxial metal sheathes and
the dielectric environment. The interaction among the propagating
modes of SPPs in these SPWs would cause the energy diffraction if no
other mechanism counteracts it, as shown below. We then start by
exploring the guiding modes in each SPW to gain some insights. By
exactly solving the Maxwell equations, we can get the dispersion
relation and field distributions
$[\boldsymbol{E^{(n)}}(r),\boldsymbol{H}^{(n)}(r)]$ of the SPP mode
in $n$-th SPW. Here we only consider the fundamental TM mode of SPP
with nonvanishing components $E_{z}^{(n)}$, $H_{\phi}^{(n)}$ and
$E_{r}^{(n)}$, which are functions of the radial coordinate $r$. The
dispersion relation of $1$-th SPW is given in Fig. 1(c), where
$\beta$ is the propagation constant of the SPP, $n_{b}$ is the
refractive index of the environment, and $k_{0}$ is the wavenumber
in vacuum. The dot-dash line labeled by $n_{b}k_{0}$ is the
dispersion of the light in the background. From Fig. 1(c) one can
see that the propagation constant $\beta$ has a strong dependence on
the frequency and the real part is always bigger than that of the
background, which is the characteristics of the SPP. Calculations
show that $\beta_n$ have a small variation for a given frequency,
especially for the SPWs with high index, which is reasonable as the
high-index SPWs can be approximated as planar waveguides with the
same width of $w$. The field distributions of the SPP modes in
different SPWs is sensitive to the frequency. For example,
$E_{z}^{(n)}$ ($n=1,2,\cdots,6,7,\cdots$) are plotted in Fig. 1(d)
for $\lambda=700$ nm and in Fig. 1(e) for $\lambda=1550$ nm. It is
clear that the interaction between adjacent modes in case of
$\lambda=700$ nm is small enough to take a nearest-neighbor
approximation while it is not in case of $\lambda=1550$ nm.

For the whole PFW, we can express the total electric field $\boldsymbol{E}%
(r,z)$ and magnetic field $\boldsymbol{H}(r,z)$ in the superposition
of the modes in all SPWs:
$E_{r}=\sum_{n}a_{n}(z)E_{r}^{(n)}e^{-i\omega t}$,
$E_{z}=\sum_{n}a_{n}(z)\dfrac{\epsilon^{(n)}}{\epsilon}E_{z}^{(n)}e^{-i\omega
t}$, and $H_{\phi}=\sum_{n}a_{n}(z)H_{\phi}^{(n)}e^{-i\omega t}$,
where $a_{n}$ is the mode amplitude of the $n$-th SPW, $\epsilon$
and $\epsilon^{(n)}$ are the dielectric constant of the PFW and of
$n$-th SPW, respectively. For simplicity all these modes of SPWs
have been normalized by their energy flow
$I_{n}=\dfrac{1}{2}\int_{S}\operatorname{Re} \left(\boldsymbol{E}
^{(n)\ast}\times\boldsymbol{H}^{(n)}\right)
\cdot\boldsymbol{e_{z}}dS$ here. Thus the equation describing the
dynamics of the mode amplitudes $a_{n}$ can be determined by the
generalized Lorentz reciprocity theorem\cite{Chuang}, which can be
written as:
\begin{equation}
i\dfrac{d}{dz}\mathbf{CA}+(\mathbf{BC}-\mathbf{K})\mathbf{A}-\mathbf{\Gamma
}|\mathbf{A}|^{2}\mathbf{A}=0,\label{eq1}
\end{equation}
where $\mathbf{A}$ is a vector given by the elements $a_{n}$
$(n=1,2,\cdots)$ and $|\mathbf{A}|^{2}$ represents a diagonal matrix
with corresponding elements $|a_n|^2$. Matrix $\mathbf{B}$ is a
diagonal matrix with the elements given by the propagation constants
$\beta_{n}$. $\mathbf{C}$, $\mathbf{K}$ and $\mathbf{\Gamma}$ are
matrices, whose elements are defined respectively by
\begin{align}
&  C_{n,n^{\prime}}\equiv\dfrac{1}{4}\int_{S}\left(  E_{r}^{(n^{\prime}%
)}H_{\phi}^{(n)}+E_{r}^{(n)}H_{\phi}^{(n^{\prime})}\right)  dS,\\
&  K_{n,n^{\prime}}\equiv\dfrac{\omega}{4}\int_{S}(\epsilon^{(n)}%
-\epsilon)\tilde{K}_{n,n^{\prime}}dS,\\
&  \Gamma_{n,n^{\prime}}\equiv\int_{S}\dfrac{-\epsilon_0\xi\epsilon_d^2%
\alpha_{n^{\prime}}\omega}{4}\tilde{K}_{n,n^{\prime}}dS,
\end{align}
where
$\tilde{K}_{n,n^{\prime}}=E_{r}^{(n^{\prime})}E_{r}^{(n)}-\dfrac
{\epsilon^{(n^{\prime})}}{\epsilon}E_{z}^{(n^{\prime})}E_{z}^{(n)}$
, $\alpha_{n}=|E_{r}^{(n)}|^{2}+|\frac{\epsilon^{(n)}}{\epsilon
}E_{z}^{(n)}|^{2}$.

In the above derivation for Eq. (\ref{eq1}), the change of relative
dielectric constant $\epsilon_d$ due to the introduced nonlinearity
in the dielectric layers is $\xi\epsilon_{d}^{2}|E|^{2}$ with the
Kerr coefficient $\xi$. Also, we have neglected the nonlinear
interaction among the SPWs, i.e. items including
$|a_{n^{\prime}}|^{2}a_{n}$ ($n\neq n^{\prime}$) are neglected. If
the distance $d$ is big enough so that the overlapping of adjacent
SPWs' fields can be ignored, the matrix $\mathbf{C}$ can also be
approximated as a diagonal matrix, as shown in Ref. \cite{Ye2010}.
But for the case of strong interaction, as the case shown in Fig.
1(e), the approximation is not acceptable. Therefore the Eq.
(\ref{eq1}) is suitable to more general cases, which can also be
written as a more simplified form
\begin{equation}
i\dfrac{d}{dz}\mathbf{A}+\mathbf{TA}+\mathbf{G}|\mathbf{A}|^{2}\mathbf{A}=0,
\label{eq2}
\end{equation}
where
$\mathbf{T}\equiv\mathbf{C}^{-1}\mathbf{B}\mathbf{C}-\mathbf{C}^{-1}\mathbf{K}$
and $\mathbf{G}\equiv-\mathbf{C}^{-1}\mathbf{\Gamma}$. This is
discrete matrix Schr\"{o}dinger equations. For the parameters we
adopted, we find that matrix $\mathbf{T}$ and $\mathbf{G}$ can be
approximated as a triple diagonal matrix and a diagonal one,
respectively. Neglected elements are at least one order of magnitude
less than others. Then it is reduced to the nonsymmetric discrete
nonlinear Schr\"{o}dinger equations with variable nonlinear
coefficients. In the equations, the matrix $\mathbf{T}$ is
determined by the interaction between the SPP mode in each SPW with
with its counterpart in the inner nearest SPW and outer one. Due to
the cylindric symmetry of the PFW, $\mathbf{T}$ is nonsymmetric
matrix, which leads to the different diffraction towards the center
or outside. The matrix $\mathbf{\Gamma}$ governs the nonlinearity,
which will balance the diffraction when solitons forms. The elements
$|G_{n}|$ ($n=1, 2, \cdots$), proved to decrease with the $n$ index,
give a centripetally increasing nonlinearity. Both these
characteristics contribute to the highly confined solitons in the
PFW.

\begin{figure}[ptb]
\includegraphics[width=8cm]{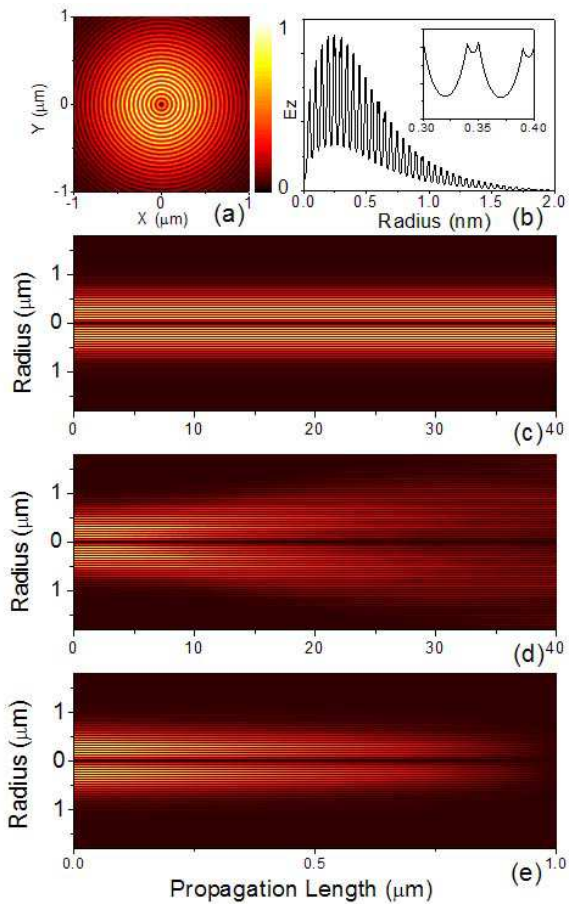}\caption{(color online)(a) the profile of the
amplitude of the longitudinal field component $E_{z}$ of the soliton
with the wavelength of $\lambda=700$ nm in the PFW. (b) the
distribution of $E_{z}$ along the radial direction, inset is an
amplification of one part for clarity. (c) The nonlinear (soliton)
and (d) the linear propagation over 40 $\mu$m distance in lossless
PFW. (e) The propagation of the soliton in the same PFW when the
loss in metal is taken into account.}%
\label{figure2}%
\end{figure}

We sought the soliton solutions in the form of
$a_n(z)=\sqrt{I_0}u_ne^{i\rho z}$, where the amplitudes $u_n$ and
$\rho$ are both independent of $z$. We also define the intensity as
$I=\sum_n|a_n|^2$ in the solution. We put our emphasis on the
unstaggered solitons\cite{Zhang2007} formed in the self-defocusing
media, with the Kerr coefficient set as $-1\times10^{-15}m^2/V^2$
and $I_0$ about $5\times10^{-4}W$. The number of layers is chosen
big enough to ensure the fields vanishing at the boundary.

\begin{figure}[ptb]
\includegraphics[width=8cm]{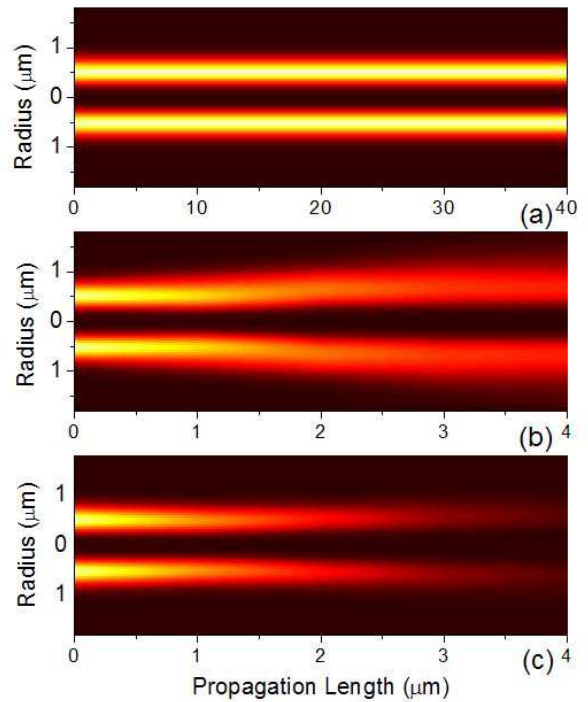}\caption{(a) The propagation of soliton
along 40 $\mu$m in the lossless PFW when the wavelength $\lambda$ is
1550nm. (b) Linear propagation along 4 $\mu$m in the same PFW. (c)
The propagation of
the soliton along 4 $\mu$m when the loss is considered.}%
\label{figure3}%
\end{figure}

Fig. 2(a) plots the profile of the amplitude of the longitudinal
field component $E_z$ of the soliton in the PFW with $I=0.05I_0$ and
Fig. 2(b), with an amplification of one part, presents the
distribution of $E_z$ along the radial direction in the case of
$\lambda=700$ nm. As $E_z$ reflects the intensity of the light, we
can conclude from Fig. 2(a) that the energy mainly concentrates
inside the region with the radius of about 400 nm. It is noteworthy
that the maximal intensity is at the circle with radius about 250 nm
other than at the center, which could be seen it more clearly from
Fig. 2(b). That is why we call it as the ring-like solitons. It is
the nonlinearity that confines the energy in a region with the size
of subwavelength, where the effect of field enhancement provided by
the SPPs is prominent, which in turn makes the nonlinearity more
achievable. The cylindric symmetry enable energy to concentrate
towards the axis, which further faciliates the confinement of
energy. The linear case where the diffraction dominates are also
plotted in Fig. 2(d) for comparison. Moreover, Fig. 2(b) also shows
that metal layers hold a large portion of the energy. For this
reason in addition to a relative big value of
$\operatorname{Im}[\beta_n]$ for visible light, the soliton
experiences a high loss in the PFW if the loss in the metal is
considered, as shown in Fig. 2(e). A gain medium is consequently
necessary for the PFW working on the visible lights.

In the case of $\lambda=1550$ nm, the ring-like solitons are also
found in the same PFW. Its propagation along 40$\mu$m in the
lossless situation with $I=80I_0$ is plotted in Fig. 2(a), while the
lossless linear propagation is in shown in Fig. 2(b). Compared with
the case of $\lambda=700$ nm, the diffraction is outstanding,
because of the remarkable overlapping between the adjacent SPWs'
fields and thus a strong interaction among the SPPs in different
SPWs, as we discussed before. And consequently the highly confined
soliton is achieved with a relative strong intensity. However, as
the portion of the field residing inside the metal layers decreasing
correspondingly, the transmission distance is enlarged to some
extent in the lossy PFW, which is clearly seen by comparing Fig.
3(c) with Fig. 2(e).

\begin{figure}[ptb]
\includegraphics[width=8cm]{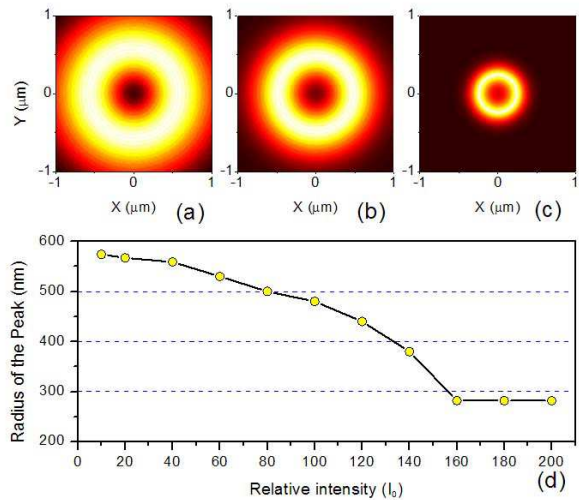}\caption{(a-c) The profiles of the
solitons with the intensity of $20I_{0}, 100I_{0}, 180I_{0}$
respectively. The values in each figure have been normalized to
their maximum. (d) Radius of the soliton's peak
changing with the intensity.}%
\label{figure4}%
\end{figure}

For the lattices solitons, increasing both the intensity and the
nonlinear coefficient will enhance its nonlinear effect and
therefore make the energy more confined. In the PFW, this
characteristics is particularly obvious. The profiles of the
solitons under three values of the intensity $I=20I_0, 100I_0,
180I_0$ are drawn in Fig. 4(a)-(c). The energy of the soliton is
confined in a ring of less and less width with the increasing of
intensity. Furthermore, because of the nonsymmetry of the matrix
$\mathbf{T}$, increasing the intensity also makes the energy moving
centripetally. It is an exciting result that the highly confined 2D
soliton in Fig. 4(c) has a transverse size far small than the
wavelength. The peaks of these solitons with more values of
intensities are plotted in Fig. 4(d). Calculation shows that when
the intensity reaches a certain value, the peak stops moving. The
reason is that in this situation the nonsymmetry of the diffraction
is trivial for the considerable nonlinearity. Besides, as the
matrixes $\mathbf{T}$ and $\mathbf{G}$ are determined by the field
distribution of SPWs, the soliton shape could be modified by
adjusting the radius and width of each metal layer.

In conclusion, we designed a PFW composed of the MDM and predicted
theoretically the form of highly confined subwavelength solitons in
the PFW with Kerr nonlinearity both for the visible light and the
near-infrared light. The equations describing the solitons have been
obtained in an compact matrix form. This kind of PFW is expected to
take the role of OFW in future in the plasmonics-based
communications and other nano-photonic applications, such as
lithography, beam shaping, {\it et al.}.

This work is supported by the National Natural Science Foundation of China
(Grant Nos. 11004015 and 61078079), National Basic Research Program of China
(Grand No. 2010CB923200), and the Doctoral Fund of Ministry of Education of
China (Grant No. 20100005120017).

*Email: jyyan@bupt.edu.cn


\begin{thebibliography}{99}                                                                                               %


\bibitem {Maier2001}S. A. Maier, M. L. Brongersma, P. G. Kik, S. Meltzer, A.
A. G. Requicha, and A. Atwater, Adv. Mater. \textbf{13}, 1501
(2001).

\bibitem {Maier2005}S. A. Maier, Current Nanoscience \textbf{1}, 17 (2005).

\bibitem {Gramotnev2010}D. K. Gramotnev, and S. I. Bozhevolnyi, Nature
Photonics \textbf{4}, 83, (2010).

\bibitem {Pitarke2007}J. M. Pitarke, V. M. Silkin, E. V. Chulkov, and P. M.
Echenique, Rep. Prog. Phys. \textbf{70}, 1 (2007).

\bibitem {Barnes2003}W. L. Barnes, A. Dereux, and T. W. Ebbesen,
\nat\textbf{424}, 824 (2003).

\bibitem {Lee2010}B. Lee, I. M. Lee, S. Kim, D. H. Oh, and L. Hesselink, J.
Mod. Opt. \textbf{57}, 1479 (2010).

\bibitem {Husakou2007}A. Husakou, and J. Herrmann, \prl\textbf{99}, 127402 (2007).

\bibitem {Vukovic2011}S. M. Vukovi\'{c}, Z. Jak\v{s}i\'{c}, I. V. Shadrivov,
Y. S. Kivshar, Appl. Phys. A \textbf{103}, 615 (2011).

\bibitem {Lin2006}C. W. Lin, K. P. Chen, C. N. Hsiao, S. Lin, C. K. Lee,
Sensors and Actuators B \textbf{113}, 169 (2006).

\bibitem {Peleg2009}O. Peleg, M. Segev, G. Bartal, D. N. Christodoulides, and
N. Moiseyev, \prl\textbf{102}, 163902 (2009).

\bibitem {Avrutsky2007}I. Avrutsky, I. Salakhutdinov, J. Elser, and V.
Podolskiy, \prb\textbf{75}, 241402(R) (2007).

\bibitem {Zhang2009}G. Bartal, G. Lerosey, and X. Zhang, \prb\textbf{79},
201103 (2009).

\bibitem {LatticeSoliton}Y. V. Kartashov, B. A. Malomed, and L. Torner,
\rmp\textbf{83}, 247 (2011).

\bibitem {Zhang2007}Y. Liu, G. Bartal, D. A. Genov, and X. Zhang, \prl\textbf{99}%
, 153901 (2007).

\bibitem {Ye2010}F. Ye, D. Mihalache, B. Hu, and N. C. Panoiu,
\prl\textbf{104}, 106802 (2010); \ol\textbf{36}, 1179 (2011).

\bibitem {Jiang2004}C. Jiang, S. Markutsya, Y. Pikus, and V. V. Tsukruk,
Nature Mater. \textbf{3}, 721 (2004).

\bibitem {Jzksic2011}Z. Jak\v{s}i\'{c}, S. M. Vukovi\'{c}, J. buha, and J.
Matovic, J. Nanophotonics \textbf{5}, 051818 (2011).

\bibitem {Fedutik2007}Y. Fedutik, V. Temnov, U. Woggon, E. Ustinovich, and M.
Artemyev, J. Am. Chem. Soc. \textbf{129}, 14939 (2007).

\bibitem {Rho2010}J. Rho, Z. Ye, Y. Xiong, X. Yin, Z. Liu, H. Choi, G. Bartal,
X. Zhang, Nature Comm. \textbf{1}, 143 (2010).

\bibitem {Johnson}P. B. Johnson and R. W. Christy, \prb\textbf{6}, 4370 (1972).

\bibitem {Smith2010}E. J. Smith, Z. Liu, Y. Mei, and O. G. Schmidt, Nano Lett.
\textbf{10}, 1 (2010).

\bibitem {Chuang}S. L. Chuang, J. Lightwave Technol. \textbf{5}, 5 (1987);
\textbf{5}, 174 (1987).
\end{thebibliography}
\end{document}